\documentstyle [11pt]{article}
\begin{document}
%
%
\newcommand{\mod}[0]{\mbox{ mod }}
\newcommand{\Abs}[1]{|#1|}
\newcommand{\Tr}[0]{\mbox{Tr}}
\newcommand{\EqRef}[1]{(\ref{eqn:#1})}
\newcommand{\FigRef}[1]{fig.~\ref{fig:#1}}
\newcommand{\Abstract}[1]{\small
   \begin{quote}
      \noindent
      {\bf Abstract - }{#1}
   \end{quote}
    }
\newcommand{\FigCap}[2]{
\ \\
   \noindent
   Figure~#1:~#2
\\
   }
\newcommand{\beq}{\begin{equation}}
\newcommand{\eeq}{\end{equation}}
\newtheorem{Th}{Theorem}
\newtheorem{Le}{Lemma}
\newenvironment{th}{\vspace{0.3cm} \em }{\vspace{0.3cm} }
%
%
%
%
\title{On the effect of pruning on the singularity structure of zeta functions}
\author{
Per Dahlqvist \\
Mechanics Department \\
Royal Institute of Technology, S-100 44 Stockholm, Sweden\\[0.5cm]
}
\date{}
\maketitle

\ \\ \ \\

%
\Abstract{
We investigate the topological zeta function for unimodal maps in general
and dynamical zeta functions for the tent map in particular.
For the generic situation, when the kneading sequence 
is aperiodic,
it is shown that the zeta functions have a natural boundary along its
radius of convergence, beyond which the function lacks analytic continuation.
We make a detailed study of the function $\prod_{n=0}^{\infty}(1-z^{2^n})$
associated with sequences of period doublings.
It is demonstrated that this function has a dense set of poles and
zeros on the unit circle, exhibiting a rich number theoretical structure.

%
}

\ \\
\section{Introduction and preliminaries}

In chaotic dynamics one aims at a global description of the phase space
flow rather than a description of individual trajectories. 
The appropriate tool
for this analysis 
is evolution operators (Ruelle-Perron-Frobenius operators).
It
requires a rather extensive
mathematical apparatus to understand the spectrum of this operator in detail.
The problem is that one has to restrict the functions space, 
acted on by the operator, in order to capture only the physically relevant
spectrum.
However it seems as if, in many cases,
this relevant spectrum can be identified
with the spectrum of poles of {\em zeta functions} or
(even better) zeros of Fredholm determinants. These functions 
require only knowledge of the periodic orbits and their invariants, such
as stability and length and no reference to specific function classes.
The link between the spectra of evolution operator and their zeta functions
has been rigorously established only for restricted classes of systems, 
like e.g. expanding (or hyperbolic) systems with a symbolic dynamics of
finite subshift type \cite{Rue,Rugh}. For such systems
the Fredholm determinants may even
be entire functions.

However,
there are many ergodic systems of physical relevance that do not fall into
this class of systems.  
We can then no longer expect singularities to be absent from the
zeta function. 
There are strong indications that for intermittent systems the zeta functions
exhibit branch point singularities \cite{Isola,PDsin,PDinter}. 
The corresponding branch cut could possibly
be associated with a continous component of the spectrum of the evolution
operator.
There is evidence that
these singularities do carry
important information about the dynamics of the system, such as 
power law decay of correlations and phase transitions among e.g. the
generalized Lyapunov exponents \cite{PDsin,PDlyap,DAcorr}.

In this paper we will study the
analytic structure of the
zeta function for expanding systems with no
symbolic dynamics of finite subshift type.

We will study the simplest conceivable system, namely one-dimensional
unimodal (one-humped) maps $x \mapsto f_\lambda(x)=\lambda\cdot g(x)$
with one external control parameter $\lambda$ \cite{MSS,Guck,CE}.

The kneading sequence $\underline{I}_\lambda$ is the orbit of the critical
point, represented by a string of $R$ and $L$'s depending on the branch
of the map visited. 
The ordering of kneading sequences with respect to the parameter 
$\lambda$ do not depend on the details of the unimodal
$g(x)$ if it has strictly negative Schwartzian derivative \cite{CE,MSS}.
For the tent map this derivative is zero and the sequences 
of period doublings
are
squeezed into a zero length intervals in parameter space.

The kneading sequence can be of one of three types \\
\indent a) It maps to the critical point again, after $n$ iterations.
If so, we adopt the convention to 
terminate the kneading sequence with a $C$,
and refer to the kneading sequence as finite.\\
\indent b) Preperiodic, i.e. it is infinite but with a periodic tail.\\
\indent c) Aperiodic.\\

Much dynamic information is encoded in 
the following one-parameter
family of zeta functions \cite{Beck}
\begin{equation}
1/\zeta(\beta,z)=\prod_p (1-\frac{z^{n_p}}{|\Lambda_p|^{1-\beta}})  \  \ .
\label{eqn:dyndef}
\end{equation}
The product in \EqRef{dyndef}
runs over all primitive periodic orbits $p$, having period
$n_p$ and
stability $\Lambda_p=\frac{df^{n_p}}{dx}|_{x=x_p}$ with $x_p$ being any
point along $p$.
We will only study a very special case of this function in detail, 
namely the
topological zeta function
\begin{equation}
1/\zeta_{top} (z) \equiv 1/\zeta(\beta=1,z)=\prod_p (1-z^{n_p}) \  \ ,
\end{equation}
and then turn to some speculations about $\beta<1$ at the end.
The leading zero  $z_0$ of $1/\zeta_{top} (z)$ 
(the one with smallest modulus) and the
topological entropy $h$ are related by $h=-\log z_0$.
There is one unimodal map for which
$1/\zeta_{top} (z)$ contains all information concerning 
$1/\zeta(\beta,z)$ for  all $\beta$,
namely the following piecewise linear map, called the
{\em tent map} \cite{Dorfle,Shi,Yosh}
\begin{equation}
x \mapsto f(x)= \left\{ \begin{array}{ll}
\lambda \cdot x & x \in [0,1/2] \\
\lambda \cdot (1-x) & x \in (1/2,1] \end{array} \right.    \label{eqn:tent}
\  \ .
\end{equation}
where the parameter $\lambda \in [0,2]$
Since $|\Lambda_p| =\lambda^{n_p}$ we have 
\begin{equation}
1/\zeta(\beta,z)=1/\zeta_{top} (z/\lambda^{1-\beta})  \  \  . 
\label{eqn:Resc} \end{equation}
The topological entropy is $h=-\log\lambda $.

The set of periodic points of the tent map is countable. A consequence of 
this fact is that the set of parameter values for which the kneading sequence
is finite or eventually periodic are countable and thus of measure zero and
consequently {\em the kneading sequence is aperiodic for almost all $\lambda$}.
For general unimodal maps the corresponding statement is that
the kneading sequence is aperiodic for almost all topological entropies.

It is, in fact, easy to write down the expanded topological
zeta function corresponding to a given kneading sequence
$\underline{I}_\lambda =PC$, of length $n$, 
where $P=s_1 s_2 \ldots s_{n-1}$ is
a string of symbols where each symbol $s_i = L \mbox { or } R$. 
Now let $a_i=1$
if $s_i=L$ and  $a_i=-1$ if $s_i=R$.
In appendix A we show the following

\begin{Th}
The expanded zeta function is a polynomial
of degree $n$ 
\begin{equation}
1/\zeta_{top} (z)=\prod_p (1-z^n_p)=(1-z)\cdot \sum_{i=0}^{n-1}b_i z^i
\  \ ,   \label{eqn:simppre}
\end{equation}
where
\begin{equation}
\begin{array}{ll}
b_0=1& \\
b_i=b_{i-1}a_i &1\leq i \leq n-1  \label{eqn:simp}
\end{array}  \  \ .
\end{equation}
If the kneading sequence is infinite (i.e. type b) or c)) the finite sum is
replaced by an infinite series.
\end{Th}

The function $\sum_{i=0}^{n-1}b_i z^i$ 
is called the characteristic polynomial in the literature
\cite{CE,Jon}. In ref. \cite{Dorfle} it appeared 
as a characteristic function whose zeros corresponds to eigenvalues
of the evolution operator acting on the space of piecewise constant functions.

An important consequence of \EqRef{simp} is that the sequence 
$\{ b_i \}$ has a periodic tail if and only 
f the the sequence $\{ a_i \}$ has
one. We will make use of this observation in section 3.

\section{A quick journey through parameter space}

\begin{table}

\begin{tabular}{|l|l|}
\hline
$I(C)$ &  $1/\zeta_{top}(z)/(1-z)$ \\
\hline
$RC$&\\
$RLRC$&\\
$RLRRRLRC$&\\
$a) H^\infty (R)$&$\prod_{n=0}^{\infty}(1-z^{2^n})$\\
$RLRRRC$&\\
$RLRRRRRC$&\\
b) $RLR^\infty$& $(1-2z^2)/(1+z)$\\
$RLRRRRRRC$&\\
$RLRRRRC$&\\
$RLRRC$&\\
$RLRRLRC$&\\
c) $RLC$& $(1-z-z^2)$\\
$RLLRLC$&\\
$RLLRLRC$& \\
$RLLRC$&\\
$RLLRRRC$&\\
$RLLRRC$&\\
$RLLRRLC$&\\
$RLLC$&\\
$RLLLRLC$&\\
$RLLLRC$&\\
$RLLLRRC$&\\
$RLLLC$&\\
$RLLLLRC$&\\
$RLLLLC$&\\
$RLLLLLC$&\\
d) $RL^\infty$& $(1-2z)/(1-z)$ \\
\hline
\end{tabular}

\caption{Ordered kneading sequences up to length seven and some longer }
\end{table}

The universal sequence of finite kneading sequences is a well studied subject
but to make ourselves acquainted with \EqRef{simp}
we will discuss some special cases . The letters
a) to d) refer to table 1.

\vspace{0.2cm}

{\bf a)} $\underline{I}_\lambda=H^\infty(R)$, i.e. the harmonic extension
of $RC$, see \cite{MSS}.
This is the accumulation point of the first cascade of period doublings. 
The topological 
zeta functions is given by
$1/\zeta_{top} (z)=(1-z)\cdot\prod_{n=0}^{\infty}(1-z^{2^n})$. 
The kneading sequence is aperiodic.
We will study this function
in some detail in the next section.
Similar cascades of period doublings accompanies all periodic orbits,
or equivalently phrased, a sequence of harmonics follows every finite kneading
sequence. 

\vspace{0.2cm}

{\bf b)} $\underline{I}_\lambda=RLR^\infty$. This is a so called
{\em band merging point}.
The inverse topological zeta function is 
$1/\zeta_{top} (z)=(1-z)(1-2z^2)/(1+z)$.
The kneading sequence is of type b).
The preperiodic kneading sequence give rise
to the denominator $(1+z)$ above and thus a pole of $1/\zeta_{top} (z)$.
The two leading zeros $z=\pm 1/\sqrt{2}$ are of the same size. 
For the tent map,
a slight
increase of the control parameter will yield 
a {\em gap} between the leading and next to leading zero
and strong mixing.
This point, and higher band merging points,
has been extensively studied in the literature \cite{Dorfle,Shi,Yosh}

\vspace{0.2cm}

{\bf c)} $\underline{I}_\lambda=RLC$.
The last Sarkovski orbit \cite{CE}
enters the system, orbits of arbitrary period
are allowed. 
The kneading sequence is $RLC$ and via eqs \EqRef{simppre} and
\EqRef{simp} we get
 $1/\zeta_{top} (z)=(1-z)\cdot (1-z-z^2)$. This can also 
 be realized as follows
(cf. refs. \cite{AAC} for details).
The kneading sequence is $RLC$, so the only forbidden subsequence
is $RLL$. All allowed periodic orbits, except $\overline{L}$, can
can be built from a alphabet with letters $\underline{RL}$
and $\underline{R}$, following ref. \cite{AAC} we write this alphabet
as $\{\underline{RL}, \underline{R} ; \overline{L}\}$,
yielding the topological zeta function 
$1/\zeta_{top} (z)=(1-z)(1-z-z^2)$.
The leading zero is the inverse golden mean $z_0=(\sqrt{5}-1)/2$.

\vspace{0.2cm}

{\bf d)} $\underline{I}_\lambda=RL^\infty$.
The symbolic dynamics is an unrestricted binary one
with alphabet $\{\underline{R}, \underline{L}\}$, 
the zeta function
is simply $1/\zeta_{top} (z)=1-2z$.
The kneading sequence is of type b) but the associated
denominator $(1-z)$ gets canceled by the prefactor in \EqRef{simppre}.

\section{Lacunary zeta functions}

Consider a function $f(z)=\sum c_i z^i$ defined by a power series.
A meromorphic functions has a well defined analytic continuation in the
entire complex plane, except at the poles.
Functions exhibiting branch point singularities 
can be continued except on
certain imposed cuts. 

But there is a class of functions that cannot be
analytically continued outside a region (typically a disk),
the boundary of this region is called
a {\em natural boundary}. 
There are some special cases where it is easy to detect
the presence of a such a boundary. 
The classical example is
{\em gap} series or {\em lacunary} series, first studied by Fredholm,
Weierstrass and Hadamard. 
The easiest example is perhaps provided by the series
$f(z)=\sum z^{2^n}$. The increasing {\em gaps} in the sequence of 
coefficients make the function noncontinuable beyond the circle
of convergence $|z|=1$.

Another example is when the coefficients can take on only a finite number of
values and the sequence of coefficients does not ultimately
become periodic, then the function again
has the unit circle as a natural boundary
\cite{Hille}. This result can be directly applied to the expansion of 
$1/\zeta_{top}$, cf eq. \EqRef{simp}, 
for an aperiodic kneading sequence.
We can thus get the following theorem for free

\begin{Th}
The topological zeta function $1/\zeta_{top}$ 
for unimodal maps has the unit circle as a natural
boundary for almost all topological entropies and for the tent map
\EqRef{tent}, for almost
all $\lambda$.
\end{Th}

Below we are going to study a specific example of zeta functions
exhibiting a natural boundary.
The concept of {\em harmonic extension}, and the
associated cascades of period doublings is a central
concept for the description of unimodal maps.
This motivates a close study of
the function
\begin{equation}
\Xi (z)=\prod_{n=0}^{\infty}(1-z^{2^n})   \label{eqn:Xi}  \  \ .
\end{equation}

The expansion of $\Xi (z)$ begins as $\Xi (z)=1-z-z^2+z^3-z^4+z^5 \ldots$.
The radius of convergence is obviously unity. The simple rule 
governing the expansion will effectively
prohibit any periodicity among the coefficients making
the unit circle a natural boundary. We will not follow this line of thought,
we will instead make a detailed
study of the fine structure of this boundary and, as a by product, 
conclude that
the unit circle is opaque.

First we rewrite $\Xi $ as
\begin{equation}
\Xi (z)=\prod_{n=0}^{\infty}(1-\chi_n)   \  \ ,
\end{equation}
where the $\chi_n$'s are defined iteratively:
\begin{equation}\begin{array}{c}
\chi_{n}=\chi_{n-1}^2\\
\chi_0=z
\end{array}\  \ .
\end{equation}
We write the $\chi_n$ on polar form $\chi_n=r_n \exp(2\pi i x_n)$
where $r_n$ and  $x_n$ are real numbers and  $x_n$ is restricted to the unit
interval.
The iteration rule is
\begin{equation}\left\{
\begin{array}{c}
r_{n}=r_{n-1}^2=r_0^{2^n}\\
x_{n}=2x_{n-1} \mod 1
\end{array}\right. \label{eqn:2rec} \  \ .
\end{equation}
Suppose now we choose $r_0$ close to, but inside the unit circle
$r_0=1-\epsilon$. Then $r_n$ will be close to unity for
$n\leq N\approx \log 2/\log \epsilon$ and then decrease to zero quickly.
By choosing $\epsilon$ 
we effectively select $N(\epsilon)$ factors of the product representation
of $\Xi$  and by decreasing $\epsilon$ we can conclude whether
$\Xi (z)$ diverges to $\infty$ or $0$ or perhaps converge
to a finite value as $z$ approach $\exp(2\pi i x)$ from within; we thus refer
to  $z=\exp(2\pi i x)$ as a pole, zero or a regular point respectively.
This reasoning is made rigorous in appendix B.

We will now focus our attention on the
second recursion relation in \EqRef{2rec} which happens to be the
binary shift map,
and we will restrict our attention to rational
$x_0$. Take $x_0=p/q$ to be proper reduced fractions, that is $q\in N$
and
\begin{equation}
p\in \Omega_q=\{p;p\in N, 1\leq p\leq q, (p,q)=1\}    \  \ ,
\end{equation}
where $(p,q)$ denote the {\em greatest common divisor} of $q$ and $p$.
The number of elements in $\Omega_q$ is Euler's function $\phi(q)$.
We are thus led to study the following map from $\Omega_q$ to itself
\begin{equation}
p_n =2 p_{n-1} \mod q \label{eqn:itself} \  \ .
\end{equation}

\begin{table}

\begin{tabular}{|l|l|l|l|l|}
\hline
$q$ & $\phi(q)$  &  $l_q$ & $F_q(1)$ &  \\
\hline
3 & 2  & 2  & 3  & $\infty$ \\
5 & 4  & 4  & 5  & $\infty$ \\
7 & 6  & 3  & 7  & $\infty^2$ \\
$9=3^2$ & 4  & 4  & 5  & $\infty$\\
11 & 10  & 10  & 11  & $\infty$ \\
13 & 12  & 12  & 13  & $\infty$ \\
$15=3\cdot5$ & 8 & 4 & 1 & $1^2$  \\
17 & 16 & 8 & 17 & 0,$\infty$ \\
19 & 18 & 18 & 19  & $\infty$ \\
$21=3\cdot 7$ & 12 & 6  & 1 & $1^2$ \\
23 & 22 & 11 & 23 & $\infty^2$ \\
25=$5^2$ & 20  & 20 &  5  & $\infty$ \\
27=$3^3$ & 18 & 18 & 3 & $\infty$ \\
29  & 28 & 28  & 29 & $\infty$ \\
31 & 30 &  5  & 31 & $0^2$,$1^2$,$\infty^2$ \\
33=$3\cdot 11$ & 20 & 10  & 1 & 0,$\infty$ \\
\hline
\end{tabular}

\caption{Some number theoretic data used to study the structure of the
unit circle for the function $\Xi(z)$. Explanations given in section 3.}
\end{table}

We will separately consider the three cases.\\
\indent i/ $q=2^m$\\
\indent ii/ $q=2^m q'$ where $(q',2)=1$\\
\indent iii/ $(q,2)=1$\\
We begin with i/.
If we set $r=1^{-}$ 
then only the first $m$ factors of
the product \EqRef{Xi} are nonzero and all the subsequent will be zero
and the corresponding point on the unit circle $z=exp(2\pi i x)$ is a zero.
These are in a sense trivial zeros and
we will eventually discover that they are not the only ones.

Case ii/ reduces to case i/ after $m$ applications of the map.

Now to the more interesting behaviour of the map \EqRef{itself}
for case iii/.
Below we are going to apply some fundamental results from number theory,
they can be found in any standard textbook. 
The text below is rather compact so 
in order to digest it could be advisable to verify table 2, 
step by step during the reading.

Now $2 \in \Omega_q$. It is well known that the set $\Omega_q$ is an abelian
group with respect to multiplication modulo $q$.
This has some important consequences. The map will move points in $\Omega_q$
along cycles, because the existence of transients would imply nonexistence
of unique inverses. Secondly, if the set $\Omega_q$ is divided into
several subcycles, all of them has the same length $l_q$, and
$l_q$ is a divisor of $\phi(q)$. 
The number $l_q$ will be the smallest solution to
\begin{equation}
2^{l_q}=1 \mod q \label{eqn:Euler}  \  \ .
\end{equation}
Due to Euler's theorem we see that $l_q=\phi(q)$ is always a solution
but it is not necessarily the smallest.
In number theory one says that $2$ is a primitive congruence root modulo
$q$ if $l_q=\phi(q)$, or equivalently, that $2$ is a generator of 
the residue class modulo $q$. 

First, if $q=p$ is a prime then  $2$ may or may not be a
generator. 
There is no other way than to check it by hand (or look it up in a table).
For $p<100$ we can use $2$ as a generator for $p=$
3,5,11,13,19,29,37,53,59,61,67 and 83 but not for
7,17,23,31,41,43,47,71,73,79,89 and 97.
If 2 generates the residue class for $q=p$ it also acts as generator
for $q=p^\alpha$ unless $2^{p-1}=1 \mod p^2$.
This is an accident which is very rare
if possible, it does not occur for any prime $p<1000$. When building up table 2
we do not have to worry about this complication.

Next suppose $q$ is composite. 
Than, there does not exist any
primitive congruence root and and $l_q<\phi(q)$.
Indeed, if $q=q_1 \cdot q_2$ and $(q_1,q_2)=1$ (and $q_1,q_2 \neq 1$)
it is fairly easy to
show that $l_q \leq \phi(q)/(\phi(q_1),\phi(q_2))\leq \phi(q)/2$.
We can be more explicit than so: decompose $q$ into prime factors
$q=p_1^{\alpha_1}p_2^{\alpha_2}\ldots p_r^{\alpha_r}$ where all $\alpha_i>0$.
If all residue classes modulo $p_i^{\alpha_i}$ can be generated by 2 then
one can show that
$l_q=\{ \phi(p_1^{\alpha_1}),\phi(p_2^{\alpha_2}),\ldots,\phi(p_r^{\alpha_r})
\}$, where $\{a,b\}$ stands for {\em least common multiple} with
obvious generalization for several arguments.

The reader can now verify the third column of table 2.

The analysis above can tell 
whether or not the set $\Omega_q$ splits up into subcycles 
under action of the map $p \mapsto 2p \mod q$.
We can also tell how many subcycles in total having period $n$.
The reason is that the binary shift map admits a binary coding of its periodic
orbits. There are
\begin{equation}
M_n=\frac{1}{n}\sum_{d|n}\mu (n/d) 2^d
\end{equation}
primitive cycles of length $n$, $\mu$ is the M\"{o}bius function,
see e.g.\ ref.\ \cite{AAC} for a derivation.
For instance, $M_3=2$ so there are two period three cycles, both corresponding
to $q=7$, cf. table 2. 
$M_4=3$ so there are three cycles of period four, corresponding
to $q=5$, $q=9$ and $q=15$.

Next we address the question if a
given (sub)cycle $c$ corresponds to a pole, a zero or a regular point of the
function $\Xi(z)$. For cycle $c$ 
we define
\begin{equation}
\psi_c=\prod_{p_j \in c} (1-e^{2\pi i \;p_j/q_c})  \label{eqn:onec} \  \ .
\end{equation}
 From the previous reasoning it is clear that if 
 $|\psi_c|<1$ then the points 
$z= \exp (2\pi i \;p_j/q_c)$ for all $p_j \in c$ are
zeros of $\Xi(z)$.  If
$\psi >1$ they are {\em poles} and if
 $\psi =1$ they are regular points. 
It is not straight forward to express 
$\psi_c$ (without numerical computation) for a given cycle but the
product over all cycles with a given $q_c=q$
\begin{equation}
\Psi_q= \prod_{q_c=q}\psi_c=
\prod_{1\leq j< q, (j,q)=1}(1-e^{2\pi i \;j/q})\equiv F_q(1)  
\label{eqn:prodc}
\end{equation}
is easily expressed.
$F_q(x)$ are the so called cyclotomic polynomials frequently studied
in number theory.
We can again use well established results \cite{arit} from number theory and
state that 

{\em If $q$ is a prime power $q=p^\alpha$ then $\Psi_q=p$, and if
$q$ contains two or more distinct prime factors then $\Psi_q=1$}.

The reader can now verify the fourth column of table 2.

We were interested in the contribution $\psi_c$ from one cycle 
and if $l_q<\phi(q)$ we are only able to compute the product $\Psi_q$ of 
contributions from all cycles corresponding to $q$.
In some cases we find that two cycles $c_1$ and $c_2$ are related by complex
conjugation (again a purely number theoretical property):
$\Psi_{c_1}=\bar{\Psi}_{c_2}$ and thus $|\Psi_{c_1}|=|\Psi_{c_2}|$.
This is the case for e.g. $q=7$.
When there is no such symmetry there seems to preference to produce
some $c$ such that $|\Psi_c|<1$ leading to zeros of $\Xi (z)$ that are surely
non trivial.

In the fifth column we indicate whether the different subcycles corresponds
to zeros, poles or regular points. 
Many entries required explicit calculations.

As a special case we can state the following

\begin{Th}
The sets of zeros {\em and} poles
of $\Xi(z)$ are dense on the unit circle. 
\end{Th}

To show that the zeros (poles) are dense it suffice to consider
points $z=\exp(2\pi i p/q)$ where $q$ is a power of two (three),
$q=2^m$ ($q=3^m$).
The function $\Xi(z)$ can thus not be continued outside the unit circle.

We will only speculate what happens for irrational $x$, that is points
that are not periodic points of the binary shift map. We know that this
map is ergodic and the invariant density is uniform. Therefore, the fact the
the following integral is zero
\begin{equation}
\int_0^1 \log (1-e^{2\pi i x})dx=0  \  \ ,
\end{equation}
leads us to believe that irrational $x$ corresponds to regular points
of $\Xi (z)$.

\section{Concluding remarks}

As we remarked before, our findings concerning the topological zeta functions
can be carried over directly to general zeta function
\begin{equation}
1/\zeta(\beta,z)=\prod_p (1-\frac{z^{n_p}}{|\Lambda_p|^{1-\beta}}) \  \ ,
\end{equation}
only for piecewise linear maps when $0\leq \beta <1$ .
However for expanding maps, the topological zeta function will,
after proper rescaling of $z$, a la eq. \EqRef{Resc}, approximate
the exact expansion of the zeta function and correspond to the
fundamental part in a cycle expansion \cite{AAC}.
It is not clear in what sense the curvature corrections of \cite{AAC} are
small in this case but
it is natural to expect the natural boundary to prevail.
However, the methods used in this paper do not immediately apply to the
non-linear case so we leave that as a speculation.

The reasoning above suggests that one could expect a boundary along
\begin{equation}
|z|=\langle |f'|^{1-\beta} \rangle
\end{equation}
when there is no finite Markov partition.
$\langle . \rangle$ is some suitable defined average, presumably geometric.

In ref. \cite{BK} it is shown that
for piecewise monotone maps $1/\zeta(z)$ is holomorphic inside
\begin{equation}
z<\lim_{n \rightarrow \infty} 
\sup ( \prod_{i=1}^{n} |f'(x_i)|)^{\frac{1-\beta}{n}}
\end{equation}
where $x_{i+1}Ê= f(x_i)$ and the supremum is taken over the initial
point $x_1$. The only requirement is that the weight, in this case
$1/|f'|^{1-\beta}$ is of finite variation. (For more recent development
on this subject 
see \cite{sharp1,sharp2,sharp3}.)
If there exists a finite Markov partition we know
that this holomorphic disk 
may be limited  by a pole of $1/\zeta(z)$.
What we suggest is that 
in the generic situation with no finite Markov partition
this guaranteed holomorphic disk
has encountered
a natural boundary.

The convergence of cycle expansion depends on the distance 
between the zero to be determined and
the nearest
singularity of $1/\zeta (z)$, zeros close to a natural boundary will be
difficult to locate accurately. 
When looking for zeros of the truncated power
series expansion of $1/\zeta (z)$
one will typically find that zeros accumulate along the
boundary as the truncation length is increased
\cite{Kai}. The situation would hardly be improved by applying some continued
fraction expansion of the zeta function.
Such an approach could be a good idea if the (inverse) zeta function is
meromorphic, 
we have encountered such a situation for the preperiodic case.
But the presence of a natural boundary would presumably lead to 
accumulation of
zeros {\em and} poles along the boundary
of the successive rational approximations of the
zeta function.


\vspace{0.5cm}

I would like to thank Viviane Baladi and Predrag Cvitanovi\'{c}
for discussions and remarks.
This work was supported by the Swedish Natural Science
Research Council (NFR) under contract no. F-FU 06420-303.


\vspace{2cm}

\section*{Appendix A}

In this appendix we will prove Theorem 1, that is, we will show
that the expansion coefficients $\{ b_i \}$ in
\begin{equation}
1/\zeta_{top}=\prod_p (1-z^n_p)=(1-z)\cdot \sum_{i=0}^{n-1}b_i z^i
\end{equation}
are given by
\begin{equation}
\begin{array}{ll}
b_0=1& \\
b_i=b_{i-1}a_i=\prod_{j=1}^{i}a_i &1\leq i \leq n-1  \  \ ,
\end{array}\label{eqn:simp2}
\end{equation}
where the coefficients $\{ a_i \}$ are obtained from the
finite kneading sequence
$\underline{I}_\lambda =PC$, of length $n$,  
where $P=s_1 s_2 \ldots s_{n-1}$, according to the rule that $a_i=1$
if $s_i=L$ and  $a_i=-1$ if $s_i=R$. 
The result follows directly from the
results of
Milnor and Thurston \cite{MT}. 
We first recall some expressions from \cite{MT}.
Let $J$ be an subinterval of the unit interval, and $f^n$ the n'th iterate
of the map. Then one defines
\begin{equation} 
\theta(x,J,f^m(x))=\left\{ \begin{array}{cc}
1 & f^m(x)\in J \mbox{ and } \frac{df^m (x)}{dx}>0 \\
-1 & f^m(x)\in J \mbox{ and } \frac{df^m (x)}{dx}<0 \\
0  & f^m(x)\not\in J \end{array} \right. \  \ .
\end{equation}
We use the convention that $f^0(x)=x$.
Let the unit interval be divided into $I_1$ and $I_2$ by the critical point
$c$ (corresponding to the symbol $C$). 
For a map with one turning point, the kneading matrix is a $1 \times 2$
matrix $(K_1,K_2)$ where
\begin{equation}
K_i=\sum_{m=0}^{\infty} 
\left( \theta(c^-,I_i,f^m)-\theta(c^+,I_i,f^m) \right)
z^m  \  \ .
\end{equation}
Perturbation of the critical point as initial value leads to
the sequence $C^\pm  P \overline{C^+ P}$ of $P$ contains an even 
number of $R$'s and $C^\pm  P \overline{C^- P}$ if it contains an odd.
We also note that if $1\leq m\leq n-1$
\begin{eqnarray}
\frac{df^m (c+)}{dx}=-b_{m-1}\\
\frac{df^m (c-)}{dx}=+b_{m-1}   \  \ .
\end{eqnarray}
A short calculation now yields
\begin{equation}
K_1+K_2=\frac{\sum_{m=1}^{n} 2b_{m-1}z^m}{1-z^n}=
2z\frac{\sum_{m=0}^{n-1} b_{m}z^m}{1-z^n}  \  \ .
\end{equation}
The matrix elements $K_1$ and $K_2$
are related by
\begin{equation}
(1-z)K_1+(1+z)K_2=0  \  \ .
\end{equation}
We can now express the kneading determinant
\begin{equation}
D(z)=\frac{K_1}{1+z}=\frac{(K_1+K_2)(1+z)}{2z(1+z)}
=\frac{\sum_{m=0}^{n-1} b_{m}z^m}{1-z^n}  \label{eqn:28}  \  \ .
\end{equation}
This determinant is related to the inverse topological zeta function
$1/\zeta_{top}$ according to \cite{MT}
\begin{equation}
1/\zeta_{top}=
D(z)(1-z^n)(1-z)=(1-z)
\sum_{m=0}^{n-1} b_{m}z^m  \label{eqn:29}
\end{equation}
as claimed.

If the kneading sequence is infinite (i.e. preperiodic or aperiodic)
then the factor $(1-z^n)$
disappears from eq. \EqRef{28} (obvious) and from eq. \EqRef{29}
(according to ref. \cite{MT}) 
and the only alteration of the final result is that
$n$ is replaced by $\infty$.


\section*{Appendix B}

Consider the function
\begin{equation}
\Xi(z=r_0 e^{2\pi i x_0})=\prod_{n=0}^{\infty}(1-r_0^{2^n}e^{2\pi i x_n})
\  \ .
\end{equation}
We are interested in the behaviour of $\Xi$ when
$r_0 \rightarrow 1^{-}$. $x_0$ is kept fixed and choosen so that
the sequence $\{ x_n \}$ is periodic:
$x_{n+m}=x_n$.
We rewrite the previous equation as
\begin{equation}
\Xi(z=r_0 e^{2\pi i x_0})=\prod_{k=0}^{\infty}f(r_0^{2^{mk}}) \  \ ,
\end{equation}
where
\begin{equation}
f(r)=\prod_{l=0}^{m-1}(1-r^{2^l}e^{2\pi i x_n}) \  \ .
\end{equation}
First we consider the case $|f(1)|>1$, 
and we will show that $|\Xi|\rightarrow
\infty$ as $r_0 \rightarrow 1^{-}$.
Obviously $f(r)$ is an entire analytic function and in particular a
continous function so that there exist some
$\tilde{r}$ such that $|f(r)|>C>1$ for all $1>r\geq \tilde{r}$.
We split up the product
\begin{equation}
\Xi(r_0 e^{2\pi i x_0})=\prod_{k=0}^{\tilde{k}-1}f(r_0^{2^{mk}})\cdot
\prod_{k=\tilde{k}}^{\infty}f(r_0^{2^{mk}})\equiv
D(\tilde{r})\cdot\prod_{k=0}^{\tilde{k}-1}f(r_0^{2^{mk}}) \  \ .
\label{eqn:hupp}
\end{equation}
We choose $r_0^{2^{m\tilde{k}}}=\tilde{r}$ so that 
$D(\tilde{r})$ is a constant
\begin{equation}
D(\tilde{r})=\prod_{k=\tilde{k}}^{\infty}f(r_0^{2^{mk}})=
\prod_{k=0}^{\infty}f(\tilde{r}^{2^{mk}})  \  \ ,
\end{equation}
depending only on our choice of $\tilde{r}$.
It is finite because $\tilde{r}<1$ and $\Xi (z)$ converges inside the
unit circle. We now have
\begin{equation}
|\Xi|>D(\tilde{r}) \cdot C^{\tilde{k}}  \  \ ,
\end{equation}
which can be made arbitrary large by choosing $\tilde{k}$ sufficiently large
and thus $r$ sufficiently close to unity.

The case when $|f(1)|<1$ is worked out in 
complete analogy it is is shown that
$|\Xi|\rightarrow
0$ as $r_0 \rightarrow 1^{-}$.

The case $f(1)=1$ requires just a little more work. First 
we rewrite eq. \EqRef{hupp}
\begin{equation}
\Xi(r_0 e^{2\pi i x_0})
=D(\tilde{r})\cdot \prod_{k=0}^{\tilde{k}-1}f(\tilde{r}^{2^{m(k-\tilde{k})}})
=D(\tilde{r})\cdot \prod_{k=1}^{\tilde{k}}f(\tilde{r}^{2^{-mk}}) \  \ .
\label{eqn:itt}
\end{equation}
We want to show that the product converges when $\tilde{k}\rightarrow \infty$,
and thus $r_0\rightarrow 1^{-}$. 
We appeal again to the analyticity of $f(r)$. Since $f(1)=1$ there exist a
constant $E$ such that 
\begin{eqnarray}
|f(r)-1|<E(1-r)  & \tilde{r} \leq r < 1  \  \ .  \label{eqn:ineq}
\end{eqnarray}
The product \EqRef{itt} converges if
\begin{equation}
\sum_{k=0}^{\infty} |f(\tilde{r}^{2^{-mk}})-1| < \infty \  \ .
\end{equation}
Using \EqRef{ineq} we get the condition
\begin{equation}
\sum_{k=0}^{\infty} |f(\tilde{r}^{2^{-mk}})-1|<E
\sum_{k=0}^{\infty} (1-\tilde{r}^{2^{-mk}})  \  \ ,
\end{equation}
which converges since
\begin{eqnarray}
(1-\tilde{r}^{2^{-mk}})=O(-\log\tilde{r} \; 2^{-mk}) & k\rightarrow \infty
\  \ .
\end{eqnarray}


\newpage

\newcommand{\PR}[1]{{Phys.\ Rep.}\/ {\bf #1}}
\newcommand{\PRL}[1]{{Phys.\ Rev.\ Lett.}\/ {\bf #1}}
\newcommand{\PRA}[1]{{Phys.\ Rev.\ A}\/ {\bf #1}}
\newcommand{\PRD}[1]{{Phys.\ Rev.\ D}\/ {\bf #1}}
\newcommand{\PRE}[1]{{Phys.\ Rev.\ E}\/ {\bf #1}}
\newcommand{\JPA}[1]{{J.\ Phys.\ A}\/ {\bf #1}}
\newcommand{\JPB}[1]{{J.\ Phys.\ B}\/ {\bf #1}}
\newcommand{\JCP}[1]{{J.\ Chem.\ Phys.}\/ {\bf #1}}
\newcommand{\JPC}[1]{{J.\ Phys.\ Chem.}\/ {\bf #1}}
\newcommand{\JMP}[1]{{J.\ Math.\ Phys.}\/ {\bf #1}}
\newcommand{\JSP}[1]{{J.\ Stat.\ Phys.}\/ {\bf #1}}
\newcommand{\AP}[1]{{Ann.\ Phys.}\/ {\bf #1}}
\newcommand{\PLB}[1]{{Phys.\ Lett.\ B}\/ {\bf #1}}
\newcommand{\PLA}[1]{{Phys.\ Lett.\ A}\/ {\bf #1}}
\newcommand{\PD}[1]{{Physica D}\/ {\bf #1}}
\newcommand{\NPB}[1]{{Nucl.\ Phys.\ B}\/ {\bf #1}}
\newcommand{\INCB}[1]{{Il Nuov.\ Cim.\ B}\/ {\bf #1}}
\newcommand{\JETP}[1]{{Sov.\ Phys.\ JETP}\/ {\bf #1}}
\newcommand{\JETPL}[1]{{JETP Lett.\ }\/ {\bf #1}}
\newcommand{\RMS}[1]{{Russ.\ Math.\ Surv.}\/ {\bf #1}}
\newcommand{\USSR}[1]{{Math.\ USSR.\ Sb.}\/ {\bf #1}}
\newcommand{\PST}[1]{{Phys.\ Scripta T}\/ {\bf #1}}
\newcommand{\CM}[1]{{Cont.\ Math.}\/ {\bf #1}}
\newcommand{\JMPA}[1]{{J.\ Math.\ Pure Appl.}\/ {\bf #1}}
\newcommand{\CMP}[1]{{Comm.\ Math.\ Phys.}\/ {\bf #1}}
\newcommand{\PRS}[1]{{Proc.\ R.\ Soc. Lond.\ A}\/ {\bf #1}}
%



\begin{thebibliography}{99}
%
{\small
\bibitem{Rue}   D.~Ruelle, Inv.\ Math.\ {\bf 34} (1976) 231. 
\bibitem{Rugh}  H.~H.~Rugh, Nonlinearity 5, 1237, (1992).  
\bibitem{Isola}  S.~Isola, {\em Dynamical zeta functions and
                 correlation
                 functions for non-uniformly hyperbolic transformations},
                 Preprint, Bologna (1995).
\bibitem{PDsin} P.~Dahlqvist, 
                Nonlinearity {\bf 8},11 (1995).
\bibitem{PDinter}P.~Dahlqvist, {\em Do zeta functions for intermittent maps
                 have branchpoints}, preprint, Stockholm (1996).
\bibitem{PDlyap} P.~Dahlqvist, {\em Lyapunov exponents and anomalous 
                 diffusion of a Lorentz gas with
                 infinite horizon using approximate zeta functions}, 
                 \JSP{}to appear.
\bibitem{DAcorr} P.~Dahlqvist and R.~Artuso, {\em On the decay of 
                 correlations
                 in Sinai billiards with infinite horizon}, submitted to
                 \PLA{}. 
\bibitem{MSS}   N.~Metropolis, M.~L.~Stein and P.~R.~Stein, J.\ Combinatorial
                Theory {\bf 15} (1973) 25.
\bibitem{Guck}  J.~Guckenheimer, Inv.\ Math.\  {\bf 39} (1977) 165.
\bibitem{CE}    P.~Collet, J.\-P.~Eckmann {\em Iterated maps on the unit
                interval as dynamical systems} (Birkh\"{a}user, Boston 1980).
\bibitem{Beck}  C.~Beck, F.~Schl\"{o}gl, {\em Thermodynamics of chaotic
                systems}, Cambridge Nonlinear Science Series 4, Cambridge
                (1993).
\bibitem{Dorfle}M.~D\"{o}rfle, \JSP{40}, (1985) 93.
\bibitem{Shi}   H.~Shigematsu, H.~Mori, T.~Yoshida and H.~Okamuto, \JSP{30}
                 (1983) 649.
\bibitem{Yosh}   T.~Yoshida, H.~Mori and H.~Shigematsu,  , \JSP{31}
                 (1983) 279.
\bibitem{Jon}    L.~Jonker, Proc.\ London Math.\ Soc.\ {\bf 39} (1979) 428.
\bibitem{Hille}  E.~Hille, {\em Analytic function theory II}  (1959).
\bibitem{AAC}   R.~Artuso, E.~Aurell and P.~Cvitanovi\'{c},
                Nonlinearity {\bf 3}, 325 and 361, (1990).
\bibitem{arit}   R.~Sivaramakrishnan, {\em Classical theory of aritmetic
                 functions}, Pure and applied mathematics no. 126,
                 Marcel Dekker, New York (1989).
\bibitem{AAC}   R.~Artuso, E.~Aurell and P.~Cvitanovi\'{c},
                Nonlinearity {\bf 3}, 325 and 361, (1990).
\bibitem{MT}     J.~Milnor and W.~Thurston, in {\em Lecture notes in
                 Mathematics}, No.~1342, J.~C.~Alexander, ed.~(1988) 465.
\bibitem{BK}     V.~Baladi and G.~Keller, \CMP{127} (1990) 459.
\bibitem{sharp1} V.~Baladi and D.~Ruelle,
                 Ergod.\ Th.\ Dynam.\ Sys.\ {\bf 14}, (1994) 621.
\bibitem{sharp2} V.~Baladi and D.~Ruelle,
                 Invent.\ Math. {\bf 123}, (1996) 553.
\bibitem{sharp3} V.~Baladi, J. Func.\
                 Anal.\ {\bf 128} (1995) 226
\bibitem{Kai}    K.~T.~Hansen, thesis, Oslo (1900).
}
\end{thebibliography}
\end{document}